\documentclass[final]{svjour3}
\usepackage{subcaption}
\usepackage[wby]{callouts}
\usepackage{url}
\usepackage {hyperref}
\usepackage{overpic}
\usepackage{pict2e} % for drawing polygons
\makeatletter
\journalname{Journal of Low Temperature Physics}
\begin{document}
\newcommand{\hdblarrow}{H\makebox[0.9ex][l]{$\downdownarrows$}-}
\title{Strategies for reducing frequency scatter in large arrays of superconducting resonators}
% Force line breaks with \\
%\input{authors}
\author{
J.~Li\textsuperscript{1},
P.~S.~Barry\textsuperscript{1},
Z.~Pan\textsuperscript{1},
C.~Albert\textsuperscript{2},
T. Cecil\textsuperscript{1},
C.~L.~Chang\textsuperscript{1,2,3}, 
K.~Dibert\textsuperscript{2},
M.~Lisovenko\textsuperscript{1},
V.~Yefremenko\textsuperscript{1},
}
% Affiliations
\institute{
\textsuperscript{1}{Argonne National Laboratory, 9700 South Cass Avenue., Argonne, IL, 60439, USA} \\
\textsuperscript{2}{Dept. of Astronomy \& Astrophysics, U. Chicago, 5640 South Ellis Avenue, Chicago, IL, 60637, USA} \\
\textsuperscript{3}{Kavli Institute for Cosmological Physics, U. Chicago, 5640 South Ellis Avenue, Chicago, IL, 60637, USA} \\
}

\authorrunning{J. Li, P. S. Barry, T. Cecil, C. L. Chang, Z. Pan }

\maketitle

\begin{abstract}
Superconducting resonators are now found in a broad range of applications that require high-fidelity measurement of low-energy signals. A common feature across almost all of these applications is the need for increased numbers of resonators to further improve sensitivity, and the ability to read out large numbers of resonators without the need for additional cryogenic complexity is a primary motivation. One of the major limitations of current resonator arrays is the observed scatter in the resonator frequencies when compared to the initial design. Here we present recent progress toward identifying one of the dominant underlying causes of resonator scatter, inductor line width fluctuation. We designed and fabricated an array of lumped-element resonators with inductor line width changing from 1.8$\mu$m to 2.2$\mu$m in step of 0.1$\mu$m defined with electron-beam lithography to probe and quantify the systematic variation of resonance frequency across a 6-inch wafer. The resonators showed a linear frequency shift of $\approx$20$M$Hz (140FWHM) and 30$M$Hz (214FWHM), respectively, as they are connected to two different capacitors. This linear relationship matches our theoretical prediction. The widely used MLA photon lithography facility for MKID fabrication has a resolution on the order of 600$n$m, which could cause frequency fluctuation on the order of 100$M$Hz or 710FWHM. 

\end{abstract}

\section{Introduction}
Superconducting microwave resonators have been extensively applied as detectors for high sensitivity sensing of excitations such as charge~\cite{schoelkopf1238,brock2021fast,tosi2019} and photons 
~\cite{cptechternach,echternach2018} due to their 
simple fabrication process and frequency domain multiplexing capability. %detection. %quantum-limited amplifier~\cite{beltran2009} to readout of superconducting qubits~\cite{blais2004} and enabling strong photon-phonon coupling~\cite{resonatorTeufel}.
Among the various types of superconducting resonator-based detectors, Microwave Kinetic Inductance Detectors (MKIDs)~\cite{mkid2003,doyle2008}, one of the most popular low-temperature superconducting resonator detectors, have
found broad applications~\cite{baselmans2008,hubmayr2015,mazin2012,moore2012,rowe2016,guo2017}. With over a decade of development effort, %MKIDs array pixels has started from 
arrays of MKIDs have evolved to thousands~\cite{eyken2015,shu2018} of detectors %pixels 
per wafer with improved %enhanced 
detector sensitivity. However, scatter in resonator frequency placement, %resonator frequency fluctuation, which The
where the fabricated resonance frequencies are shifted from their designed values, has become a limiting issue for large array productivity. Various parameters could contribute to this fluctuation such as non-uniformity in the critical temperature ($T_c$) and film thickness, non-uniformity in lithographic processing, and non-uniform etch profiles across the wafer.  

One solution for mitigating frequency fluctuations is capacitor trimming~\cite{liu2017}, a method that improves the MKID frequency distribution at the cost of introducing additional steps. %This method could improve the frequency fluctuation but requires multiple steps. 
Another approach is to improve MKID materials and fabrication to improve resonator frequency placement, though this approach requires detailed studies %Till now no investigation has been carried 
to understand the fundamental parameters that result in scatter of the resonator frequency. This paper is an initial investigation along this direction, and studies how geometric variations in resonator fabrication contribute to frequency scatter. %causes the frequency fluctuation. 
Our work is motivated in part by results from our recent MKIDs fabrication where we observed significant line width variation %fluctuation 
of up to 2~um for 4~um-wide features across a 6 inch wafer. % in the photo lithography process with MLA (Maskless Laser Aligner). 
This line width variation would significantly change the inductance of our lumped element inductor since the inductance is a strong function of line width. %, especially for inductors with narrow width. 
To investigate the impact of the line width variation on frequency placement, %change on fluctuation 
we carried out a study using electron beam (e-beam) lithography to fabricate resonators with line widths varying in steps of 100~nm, and evaluated their resonance frequency placement as a function of the line width.

%\section{inductance of meander inductor}
\section{Modeling the meander inductance}
To develop some intuition about the relationship between changes in linewidth and the resonator frequency, we use 
%gain guide line about meander inductor width change on frequency fluctuation we borrowed the 
an equation for meander inductance on PCB~\cite{stojanovic2004}:
\begin{eqnarray}
%L_{mon} &=& 0.00266\cdot a^{0.0603}\cdot h^{0.4429}\cdot N^{0.954}\cdot d^{0.606} \cdot w^{-0.173}
L_{m} &=& \beta \cdot w^{\alpha}
\label{meaindeq}
\end{eqnarray}
with $w$ the width of the meander strip. Inductance change $\Delta L_{m}$ due to small line width change $\Delta w$ can be approximated by Taylor expansion around $w$: 
%(ref: https://www.allaboutcircuits.com/tools/microstrip-inductance-calculator/) \\
%where the definition of the parameters are shown in Fig.~\ref{meanderdef} with $N$ corresponding to the number of meanders.

%\begin{figure}[htbp]
%\begin{center}
%\includegraphics[width=0.4\linewidth, keepaspectratio]{equationParameters.png}
%\caption{Illustration showing the geometric parameters used in our calculation of meander inductance (see eqn.~\ref{meaindeq}).}% used in equation~\ref{meaindeq} }
%\label{meanderdef}
%\end{center}
%\end{figure}

%For small variations, $\Delta w$, in the line width, $w$, the above equation can be approximated as:
%\begin{eqnarray}
%L_{mon} &=& 0.00266\cdot a^{0.0603}\cdot h^{0.4429}\cdot N^{0.954}\cdot d^{0.606} \cdot (w+\Delta w)^{-0.173} \\
%&\approx& 0.00266\cdot a^{0.0603}\cdot h^{0.4429}\cdot N^{0.954}\cdot d^{0.606} \cdot (w^{-0.173}-0.173w^{-1.173}\Delta w).
%\label{meaindeqsimp}
%\end{eqnarray}
%We can then calculate the change in inductance as
%\begin{eqnarray}
%\Delta L_{mon} &=& 0.00266\cdot a^{0.0603}\cdot h^{0.4429}\cdot N^{0.954}\cdot d^{0.606} \cdot -0.173 w^{-1.173}\Delta w,
%\label{meaindeqsimpdel}
%\end{eqnarray}
%and the corresponding fractional change in inducatnce as
\begin{eqnarray}
\Delta L_{m}/L_{m} &=& \alpha\Delta w/w.
\label{meaindeqsimpdelratio}
\end{eqnarray}
Similarly the kinetic inductance change $\Delta L_{K}$ due to line width variation $\Delta w$ is:
\begin{eqnarray}
\frac{\Delta L_{K}}{L_{K}} &=& -\frac{\Delta w}{w},
\end{eqnarray}
where we have used kinetic inductance equation from reference \cite{anthony2010}. \\
The corresponding shift in resonator frequency $\Delta f_{o}$ is then %induced by the inductance change can be calculated as:
\begin{eqnarray}
%f_{o} &=& \frac{1}{2\pi\sqrt{LC}} \\
f_{o} + \Delta f_{o} &=& \frac{1/2\pi}{\sqrt{(L+\Delta L)C}} \\
&=& \frac{1/2\pi}{\sqrt{LC}}\left(1-\frac{1}{2}\frac{\Delta L}{L}\right) \\
&=& f_{o} -\frac{1}{2}f_{o}\frac{\Delta L}{L} \\   %-\frac{1/2\pi}{2}f_{o}\frac{\Delta L}{L} \\
\textrm{and fractional frequency shift} \nonumber \\
\frac{\Delta f_{o}}{f_{o}} &=& -\frac{1}{2}\frac{\Delta L}{L} \\ %-\frac{1/2\pi}{2}\frac{\Delta L}{L} 
&=& \frac{1}{2}\left(\alpha \frac{L_{m}}{L_{K}+L_{m}}+\frac{L_{K}}{L_{K}+L_{m}}\right)\frac{\Delta w}{w} %\frac{1}{2\pi}
\label{delfdelw}
\end{eqnarray}

\begin{figure}[h!]
\centering
\includegraphics[width=0.7\textwidth]{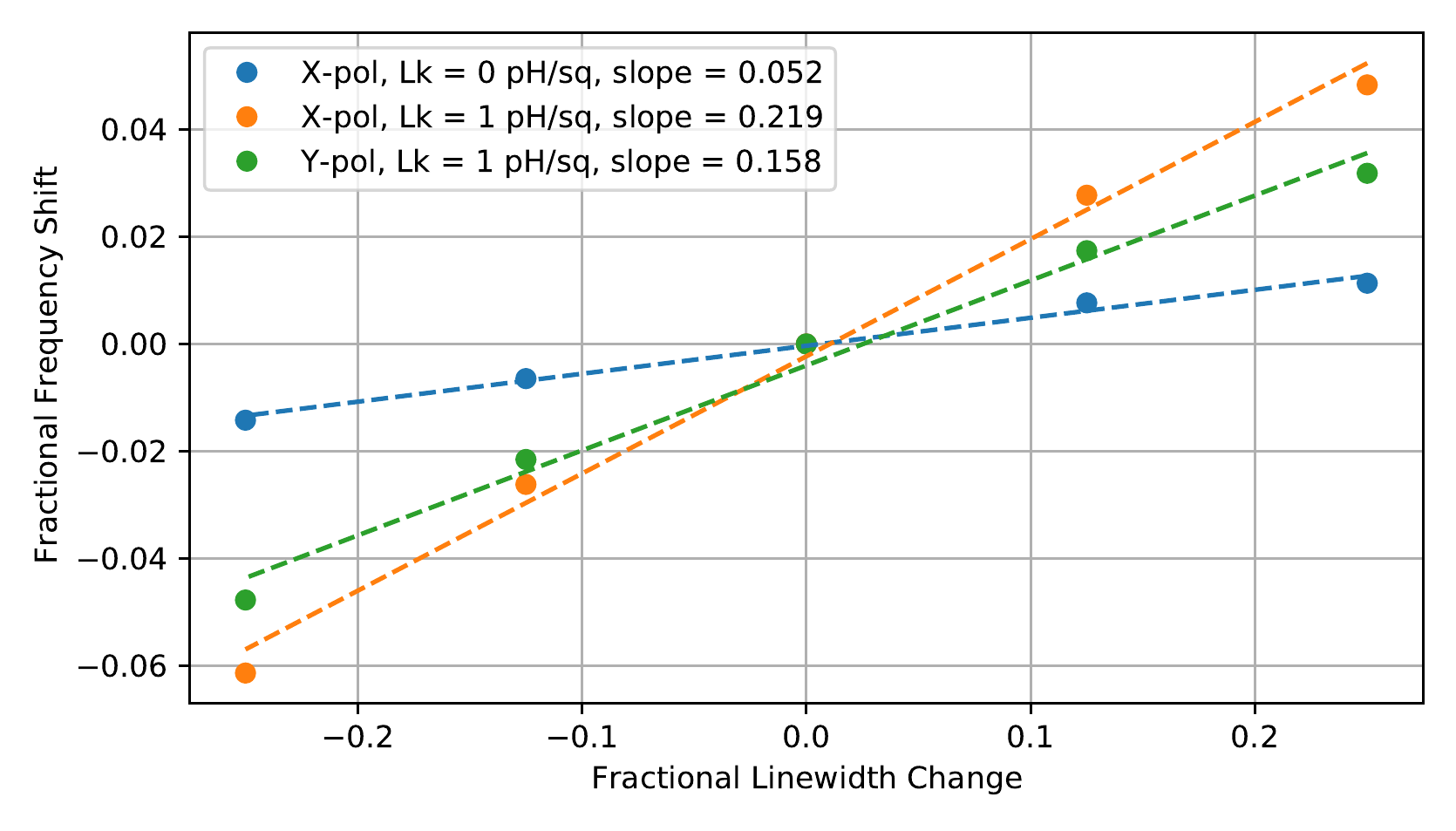}
\caption{Simulated fractional resonant frequency shift $\Delta f_{o}/f_{o}$ versus fractional line width variation $\Delta w/w$ for ``X-pol'' and ``Y-pol'' resonators. Inductor line widths were varied from 1.5 to 2.5~$\mu$m in steps of 0.25~$\mu$m. $L_{K}$ is the kinetic inductance of the superconducting thin film in the simulation.}
\label{freShiftSonnet}
\end{figure}

We carry out a more detailed model of our resonators by using a Sonnet simulation of our lumped element resonators where we change the inductor width from 1.5 to 2.5~$\mu$m in steps of 0.5~$\mu$m. Figure \ref{freShiftSonnet} shows the results of our simulations for each of our two circuit capacitor designs (see Sec.~\ref{sec:design}). %, and the results of our simulation are shown in .

\section{Test resonator design}
\label{sec:design}
\begin{figure}[h!]
     \begin{overpic}[abs,unit=1pt,scale=.3,width=0.8\textwidth]{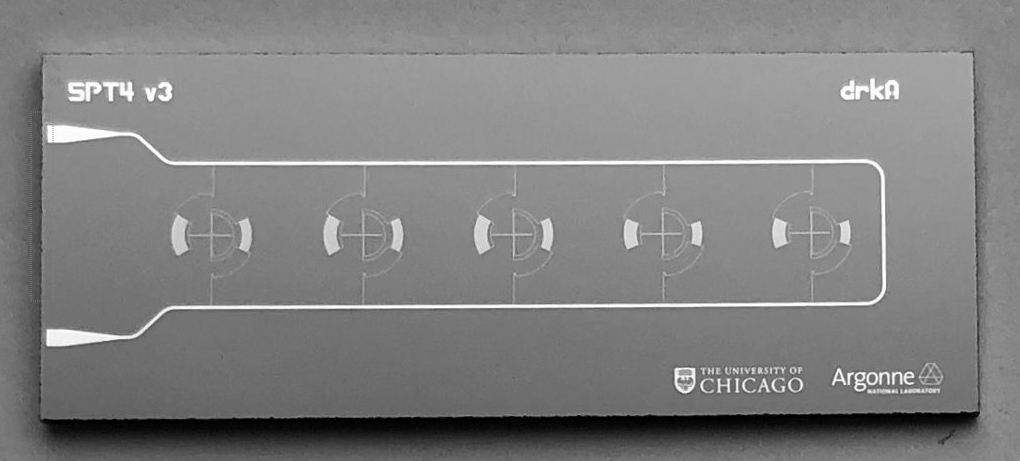}
     \put(11,14){\color{white}\large{(a)}}
     \put(140,-100){\color{black}%
     \frame{\includegraphics[scale=.06]{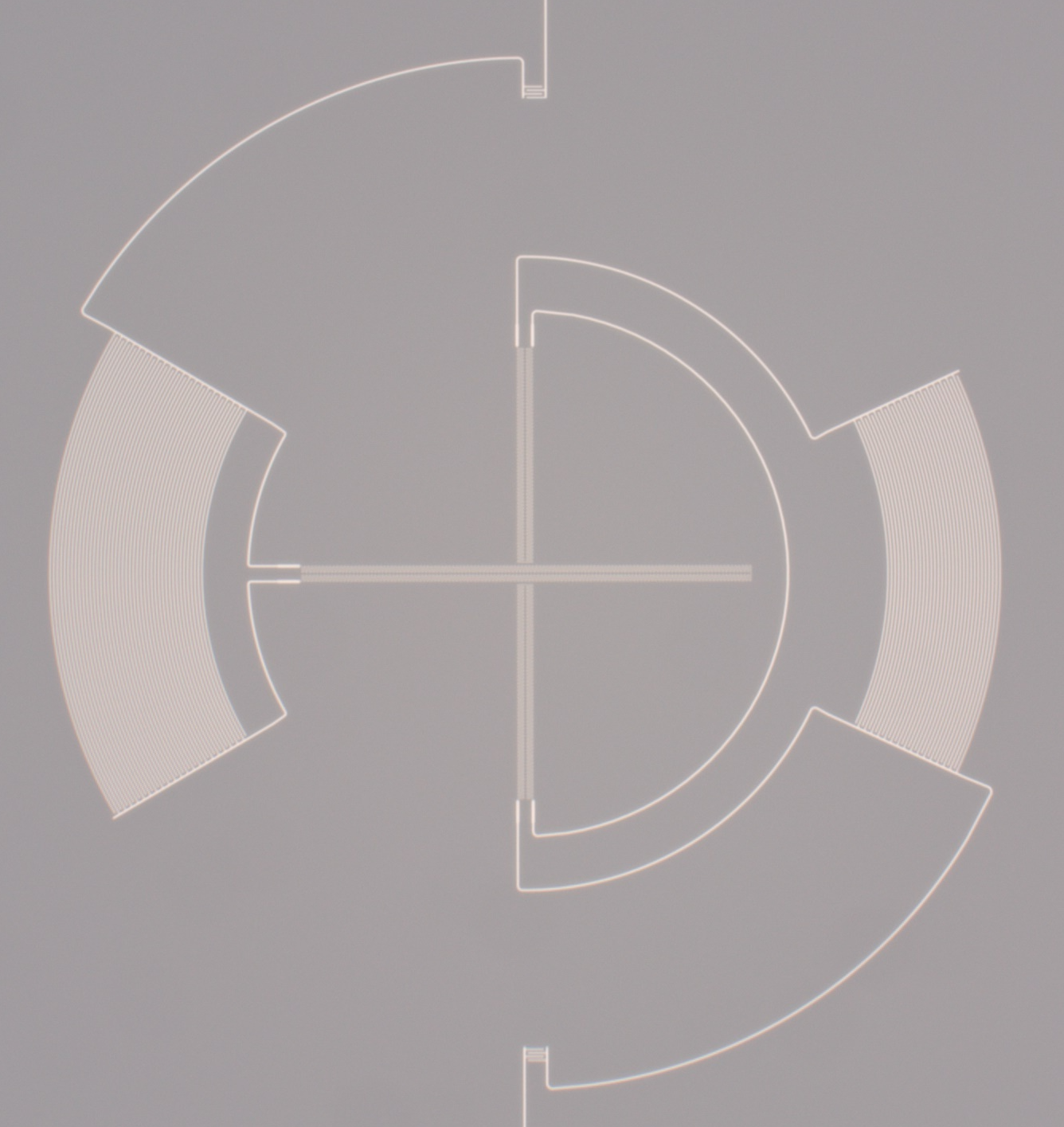}}}
     \put(150,-90){\color{white}\large{(c)}}
     \put(155,-40){\color{white}\large{$C_{x}$}}
     \put(200,-40){\color{white}\large{$C_{y}$}}
     \put(162,-65){\color{white}\small{x-pol}} 
     \put(183,-50){\color{white}\small{y-pol}} 
     \put(10,-100){\color{black}%
     \frame{\includegraphics[scale=.285]{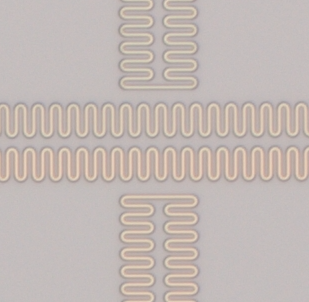}}}
     \put(20,-88){\color{white}\large{(b)}}
     \put(250,-100){\color{black}%
     \frame{\includegraphics[scale=.25]{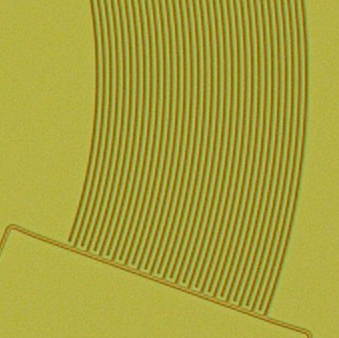}}}
     \put(260,-88){\color{white}\large{(d)}}
     \end{overpic}\
     \vskip 100pt
     \caption{\footnotesize{Images of the designed chip and components. \textbf{(a)} image of chip A which has five pixels and each pixel has two resonator that we name them Xand Y  because of the orthogonal directional of the two inductors. \textbf{(b)} zoom-in of the meander inductors. \textbf{(c)} zoom-in of the middle pixel. Left inter digital capacitor $C_{x}$ is connected to the horizontal (X) meander inductor and the right inter digital capacitor $C_{y}$ is connected to the vertical (Y) meander inductor. \textbf{(d)} zoom in of inter digital capacitor $C_{y}$.}}
\label{resodesign}
\end{figure}
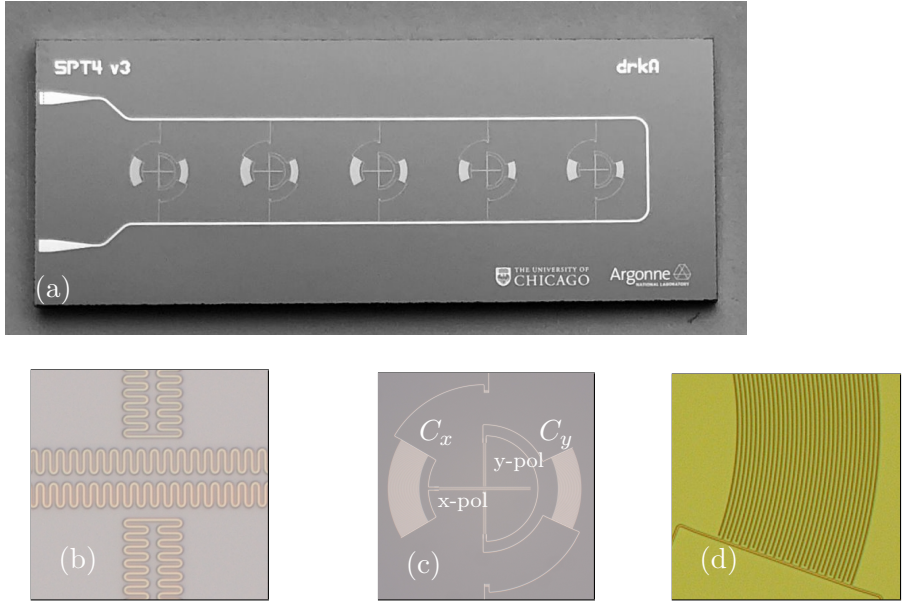

To empirically study the impact of resonator geometry on frequency placement, we designed a test device comprised of five nearly identical lumped element resonators with varying inductor linewidth. 
%The test resonators are designed with lumped inter digital capacitor and meander inductor. 
In our design (see Fig.~\ref{resodesign}), we have five pixels, with each pixel having one ``X-pol'' resonator and a ``Y-pol'' resonator. All of the ``X-pol'' (``Y-pol'') resonators have the same circuit capacitor design, $C_x$ ($C_y$), and the inductor width for both resonators is varied from 1.8 to 2.2$\mu$m in steps of 0.1$~\mu$m across the five pixels. With this configuration we expect two groups of resonances corresponding to the two circuit capacitor designs, $C_{x}$ and $C_{y}$. Within each group, we expect to have five linearly distributed resonances corresponding to the five different inductor widths.
%resonators are grouped together as a pixel which is shown in figure (\ref{resodesign}) the X and Y inductors have same line width but connected to different capacitors ($C_{x}, C_{y}$). Each resonator is connected to the measurement feed line through a inter digital coupling capacitor. $C_{x}$ and $C_{y}$ are same among the five pixels. The inductors width are varied from 1.8 to 2.2$\mu$m in step of 0.1$\mu$m across the five pixels. In the test result we would expect two groups of resonances that is determined by $C_{x}$ and $C_{y}$. Within each group there should be five ilnearly distributed resonances due to five different inductor widths.  

\vskip -150pt
\section{Fabrication}
Our devices were fabricated on a single silicon wafer, which minimizes the impact from comparing materials from different depositions. To achieve the required 100~nm linewidth resolution, the resonators were patterned using ebeam writing with the highest resolution of 10~nm. %To rule out nonuniformaity from other parameters all the resonators are on a single silicon wafer. 
Details of the fabrication process are:

The bare wafer was vacuum baked in $120^{\circ}C$ for six minutes for removing moisture. Ebeam resist of PMMA 950 A4 was spincoated at a speed of 4000rpm for 40 seconds and then baked at 180$\cdot$ for 4 minutes afterward. The inductor, capacitor, and connection lines are on a single layer and were written with a dose of 720$\mu$C/$\textrm{cm}^{2}$. The feed-through lines as well as the contact pads are on another layer and written with a dose of 800$\mu$C/$\textrm{cm}^{2}$. Proximity effect correction effectively lowers the dose for large structures, and the dose is increased purposely to compensate for it. After pattern writing, the wafer was developed with MIBK and IPA with a ratio of 1:3 for 40 seconds and rinsed in IPA for 10 seconds.

After development a 20 seconds oxygen descum process was applied to remove leftover ebeam resist. Then, a 30~nm thick Al film  was deposited via DC magnetron sputterring at a base pressure of 1.7E-8mBar. The final processing step was an overnight soak in acetone for lift-off.

We chose the two tested chips near the center of the wafer to minimize film thickness non-uniformity across the two chips. During fabrication, one of the capacitors on one of our devices (chip B) had a liftoff issue and was discarded from our measurements. This only impacted one single resonator.  %Since the chips were deposited together on the same wafer at the same time we rule out the possibility of Tc nonuniform between the two chips.

%Chips were achieved after overnight lift off with acetone. However one capacitor of chip B had significant liftoff issue and the resonator frequency was throw off from the rest of the resonators in the test result.

\section{Test results}
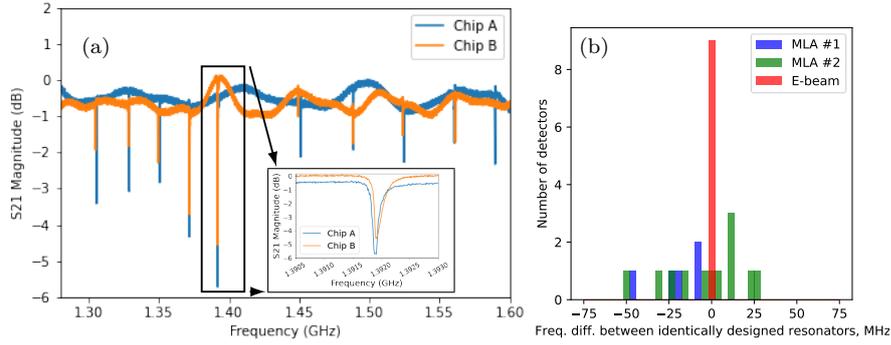
\begin{figure}
\begin{subfigure}[t]{0.52\textwidth}
\begin{overpic}[abs,unit=1pt,scale=.5]{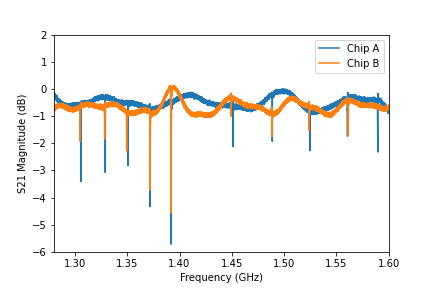}
%\put(5,45){\color{blue}\huge\LaTeX}
\put(35,110){\color{black}(a)}
\put(80,20){\linethickness{0.25mm}\color{black}\polygon(0,0)(16,0)(16,85)(0,85)}
\put(105,20){\color{black}%
\frame{\includegraphics[scale=.16]{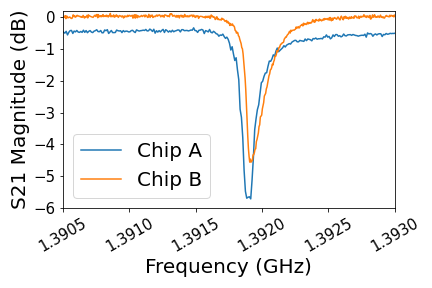}}}
\put(98,105){\linethickness{0.25mm}\color{black}\vector(1,-3.8){10}}%
\put(98,20){\linethickness{0.25mm}\color{black}\vector(1,0){6}}
\end{overpic}
\end{subfigure}
\hfill
\begin{subfigure}[t]{0.42\textwidth}
\begin{overpic}[abs,unit=1pt,scale=.47]{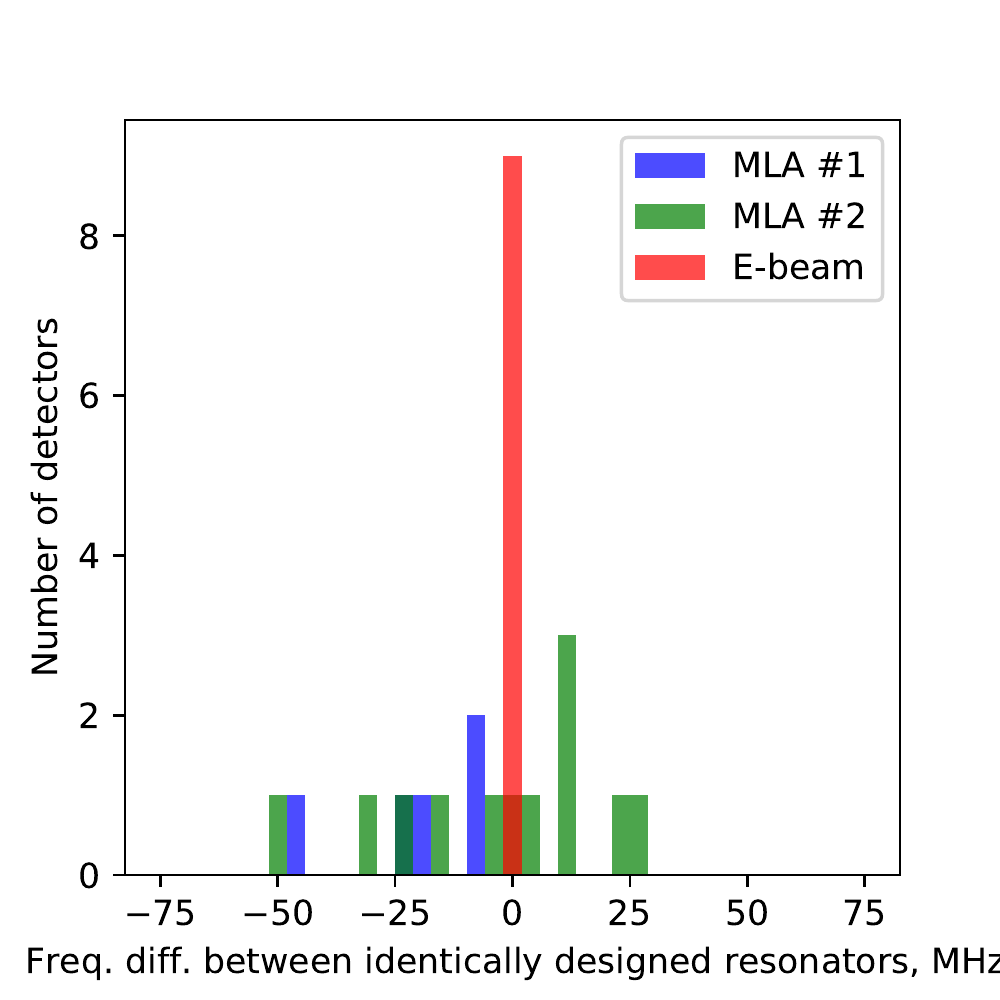}
%\put(5,45){\color{blue}\huge\LaTeX}
\put(20,110){\color{black}(b)}%
%\frame{\includegraphics[scale=.15]{ebeamcomparezoomed.png}}}
\end{overpic}
\end{subfigure}
\caption{\textbf{(a)} Tested resonator frequencies as a function of inductor line width change. Blue traces is for chip A, and orange trace is for chip B. The inset is zoomed in to the two resonances from chips A and B around 1.3917GHz. An offset of 15~KHz is observed. \textbf{(b)} Resonance scatter estimated by comparing the measured frequency for identically designed resonators. Devices fabricated using ebeam lithography with lift-off have scatter $<$1~MHz, consistent with the 10~nm resolution and our measured geometric dependence. Two other chips (MLA\#1 and MLA\#2) that were fabricated using Direct Write lithography and wet etching exhibit much larger frequency scatter.}%between chips for resonators fabricated with both ebeam and MLA proce. }
\label{testres}
\end{figure}
%The chips are tested 
We tested two chips (chip A and chip B) in a dilution refrigerator with a base temperature of 10~mK. The input lines have $\approx$70dB attenuation. The output signal is amplified with a HEMT amplifier with a gain of around 40~dB. A superconducting coaxial cable is used between the output of the sample box and the HEMT amplifier to minimize signal loss. The resonances are scanned with a standard $S_{21}$ measurement of VNA.  

\begin{figure}
\centering
\includegraphics[width=0.8\linewidth, keepaspectratio]{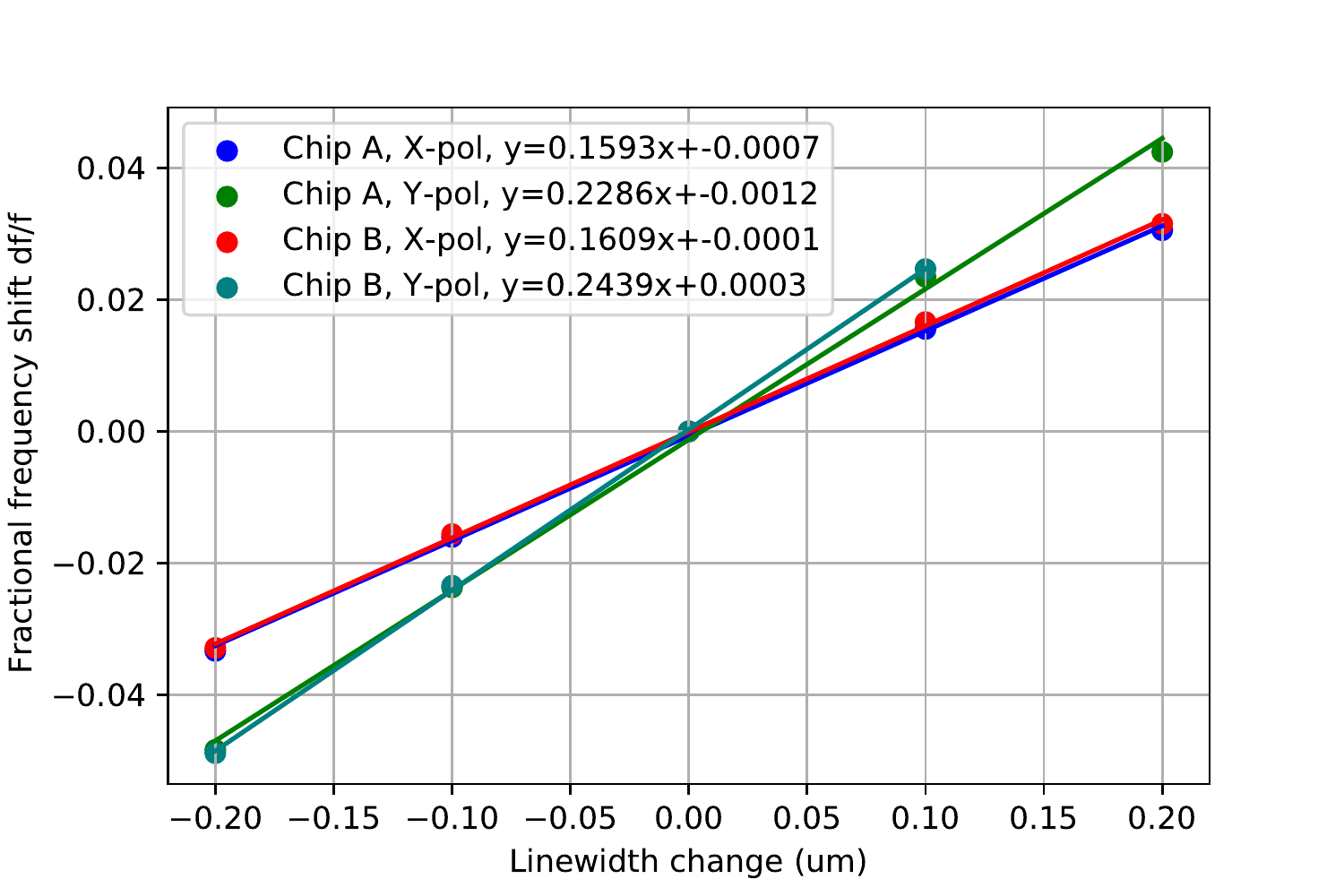}
\caption{\footnotesize{$\Delta f_{o}/f_{o}$ vs. $\Delta w$ for all the resonators of both chip A and B. ``X-pol'' stands for resonators that use the $C_{x}$ capacitor design and ``Y-pol'' stands for resonators using the $C_{y}$} capacitor design.}
\label{linearplot}
\end{figure}

Figure~\ref{testres}(a) shows the measured resonances for both chip A and B. The trace for chip A shows ten resonances with the lower (higher) frequency resonances corresponding to the resonators with the ``Y-pol'' (``X-pol'') capacitor design. %the left five are all connected to capacitors that share the same value, and the right five are connected to capacitors that share another value. The difference in the frequency step sizes among the first five and second five is due to the different capacitors used. 
Within each group, the resonances differ due to the varying inductor linewidth. In Fig.~\ref{linearplot}, we plot the measured fractional frequency shift, $\Delta f_{o}/f_{o}$, versus the change in inductor linewidth, $\Delta w$. Our data exhibits a clear linear relationship, as predicted by Eq.(\ref{delfdelw}), with a slope that is consistent with predictions from our simulations. 
%linearly due to the inductor line width change. This linear relationship between the frequency shift and inductor line width change agrees well with our theoretical prediction. Resonances at a higher frequency is shifted further than resonances at a lower frequency, which agrees with Eq. (\ref{delfdelw}). 

We estimate the frequency scatter from our ebeam process by comparing the frequencies for identically designed resonators. We find the largest scatter is $\sim$1~MHz, which is consistent with the 10~nm ebeam resolution and our measured dependence of inductance on linewidth. To give a sense of the importance of linewidth control, we compare our results with a similar measurement of frequency scatter for two other sets of resonators that were patterned using Direct Write optical lithography and wet etching (see Fig.~\ref{testres}).
%the corresponding resonance between chips A and B, and the largest scatter is about 1MHz

%To illustrate the frequency fluctuation from MLA, we compare the frequency scatter of chips fabricated with ebeam and MLA in Figure ~\ref{testres}(b). Scatter from the ebeam process is derived by comparing the corresponding resonance between chips A and B, and the largest scatter is about 3MHz. MLA 1\#1 and \#2 Resonators have different designs compared to the ebeam ones, but pairs of them are fabricated on the same wafer, and the scatter is derived by comparing the pair of chips from the same wafer. It is worth pointing out that the resonators patterned with ebeam were developed with the liftoff process while the MLA ones are through the wet etch process. 

%Figure \ref{linearplot} is the plot of $\Delta f_{o}/f_{o}$ vs. $\Delta w/w$ for the two groups of resonators of both chips A and B. All four groups show a clear linear relationship as predicted by Eq.(\ref{delfdelw}). However, the slope of the linear fit should all be close to 0.173/2 as predicted by the equation, which is not the case here. This discrepancy could be from that the equation is for a meander inductor on PCB instead of a silicon wafer, or certain parameters are frequency-dependent and can not be treated as a constant.

\section{conclusion}
We have %successfully 
demonstrated the impact of meander linewidth variation on the frequency scatter of lump element resonators. Our results are in good agreement with our simulations and show that control of resonator geometry can be a significant contributor to frequency scatter in large arrays of MKIDs.
%Linewidth scatters due to the low resolution of the photon lithography process are one of the fundamental sources of frequency scatter for large arrays of mKIDS or superconducting resonator detectors. 

\begin{acknowledgements}
Work at Argonne, including use of the Center for Nanoscale Materials, an Office of Science user facility, was supported by the U.S. Department of Energy, Office of Science, Office of Basic Energy Sciences and Office of High Energy Physics, under Contract No. DE-AC02-06CH11357. Zhaodi Pan is supported by ANL under award LDRD-2021-0186
\end{acknowledgements}

%\section{ebeam patterning process}
%\label{ebeampro}
%\nocite{*}
\bibliographystyle{unsrt}
\bibliography{aipsamp.bib}
%\bibliography{aipsamp}% Produces the bibliography via BibTeX.

\end{document}